\documentclass[prc,aps,reprint,amsmath]{revtex4-2}

\usepackage{graphicx}
\usepackage{bm}
\usepackage{array}
\usepackage{subcaption}
\usepackage{amsmath,amssymb}
\usepackage{braket}
\usepackage[breaklinks, colorlinks=true]{hyperref}

\newcommand{\unit}[1]{\,\mathrm{#1}}
\newcommand{\nl}{\nonumber\\}

\begin{document}

\title{Fermi Operator Expansion for the Hartree-Fock-Bogoliubov Theory}
\date{\today}
\author{Chengpeng Yu}
\email{yu.chengpeng@nucl.ph.tsukuba.ac.jp}
\affiliation{Center for Computational Sciences, University of Tsukuba, Tsukuba 305-8577, Japan}
\author{Takashi Nakatsukasa}
\email{nakatsukasa@nucl.ph.tsukuba.ac.jp}
\affiliation{Center for Computational Sciences, University of Tsukuba, Tsukuba 305-8577, Japan}
\affiliation{Faculty of Pure and Applied Sciences, University of Tsukuba, Tsukuba 305-8571, Japan}
\begin{abstract}
\begin{description}
    \item[Background] A variety of phases in the inner crust of neutron stars are crucial for understanding the pulsar phenomena. However, the three-dimensional coordinate-space calculation of the phases is computationally demanding.
    \item[Purpose] We aim to generalize the Fermi operator expansion (FOE) method that is effective for finite-temperature coordinate-space simulation, from the Hartree-Fock theory to Hartree-Fock-Bogoliubov (HFB) theory including the pairing effects.
    Furthermore, the periodic structure with free neutrons in the inner crust requires us to treat the system with the band theory.
    \item[Method] We give a concise proof that the generalized density matrix in the HFB theory can be obtained with the FOE.
    The Chebyshev polynomial expansion is used for calculations of the HFB band theory.
    \item[Results] Using a model for a slab phase of the inner crust,
    the FOE method produces results in good agreement with those
    based on the diagonalization of the HFB Hamiltonian.
    \item[Conclusions] The FOE method for the HFB band theory is a powerful tool for studying the nontrivial exotic structures in neutron stars.
    The FOE method is suitable for parallelization and further acceleration is possible with nearsightedness.
\end{description}
\end{abstract}
\maketitle

\section{Introduction}

Neutron stars, compact high-density celestial bodies composed of nuclear matter, exhibit layered structures
\cite{Chamel:2008ca}:
In the inner crust region, nuclei are placed to form a crystal structure with a degenerate Fermi gas of electrons and dripped free neutrons.
In the core region, the crystal structure disappears and changes into uniform nuclear matter.
Around the border region between the inner crust and the core, nuclear matter forms exotic inhomogeneous phases with nonspherical symmetry, referred to as the pasta phases \cite{RPW83,HSY84,Oya93}.
The typical pasta phases include the slab phase
and the rod phase.
The pasta phases have significant influences on various phenomena of neutron stars, such as pulsar glitches \cite{Reichley:1969, Link:1999ca, Andersson:2012, Chamel:2012zn} and magnetic field decay \cite{Pons:2013nea,Horowitz:2014xca}.

For studies of the pasta phases,
it is desirable to perform three-dimensional (3D) coordinate-space
simulations of nuclear matter based on nuclear energy density functionals (EDF)
\cite{Newton09}.
The EDF approaches normally involve
(1) solving the Kohn-Sham (KS) equations in the coordinate space by diagonalizing the KS Hamiltonian,
(2) calculating the densities of baryons based on the solution
of the KS orbitals,
(3) updating the KS Hamiltonian according to the EDF with the new densities, and
(4) repeating procedures 1--3 until the densities converge.
To treat the transport properties of the dripped free nucleons,
the method must be combined with band theory
\cite{Kashiwaba:2019ioe, Sekizawa2022, Almirante:2023gme, Yoshimura:2023hyp}.
The 3D calculation is preferable to
discover new exotic structures without assuming any spatial symmetry \cite{Kashiwaba:2020iiv, Nakatsukasa:2022hpp}.
It is also useful for studying the transition from the inhomogeneous phase
to the uniform phase of nuclear matter.

Despite the powerful and useful nature of the method,
there is a drawback in the computational demands:
With the number of lattice sites $N$ in the coordinate space,
the computational task increases as $O(N^3)$.
Here, we assume that, in each iteration, one solves the KS equations by diagonalization.
Thus, for the 3D calculations of the inner crust of neutron stars,
the computation becomes formidably expensive.
A possible way to avoid the difficulty is to introduce additional symmetries, for example, restricting the system to be practically one dimensional (1D) 
with the translational symmetry in the transverse directions
\cite{Kashiwaba:2019ioe, Sekizawa2022, Yoshimura:2023hyp, Almirante:2023gme}.
For the two-dimensional (2D) and 3D structures, the Thomas-Fermi approximation is
often adopted, a semiclassical approximation that describes the baryons solely through their local densities \cite{Xia:2022rfc, Okamoto:2013tja, Caplan:2016uvu}.
This approach significantly reduces the computation time; however,
the shell effect must be treated separately.
In addition, there is the Wigner-Seitz approximation \cite{NV73}.
It replaces a 3D periodic system with a spherical cell and uses
different boundary conditions for even- and odd-parity levels.
However, there is an ambiguity related to the boundary condition
\cite{Baldo:2006jr, Chamel:2008ca, Togashi:2017mjp, Chamel:2007it}.
Finally, Ref.~\cite{JBRW17} uses the shifted Krylov subspace method to calculate
the Green's function and the densities in the 3D coordinate space.
The method has been extended to finite temperature \cite{Kashiwaba:2020iiv}.
The performance of these methods depends on the convergence of the shifted Krylov subspace method for solution of linear algebraic equations.

In this paper, we adopt another method to accelerate the coordinate-space simulation: Fermi operator expansion (FOE) \cite{Wu:2002}.
The key idea is to relate the one-body density matrix of baryons to $f(H)$, 
where $H$ is the Hamiltonian, and $f(x)$ is the Fermi-Dirac 
distribution function at finite temperature.
One can compute $f(H)$ by expanding it into the polynomial series of $H$, and then directly obtain the densities without solving any equation.
Historically, this method was first used in condensed matter physics \cite{Goedecker:1994, Goedecker:1995, Kohn:1995}, and was known to be a method of order-$N$ $[O(N)]$ complexity \cite{Wu:2002}.
The FOE was applied to nuclear physics for the first time
with the 
3D coordinate-space calculation of finite nuclei and nuclear matter
at finite temperature \cite{Nakatsukasa:2022hpp}.
However, the work of Ref.~\cite{Nakatsukasa:2022hpp} is based on the KS scheme, and did not incorporate the pair density.
It also uses a simple periodic boundary condition for the 3D nuclear matter,
which corresponds to the crudest approximation in the band calculation.
It is important for the scattering of the free neutrons on the periodic potential in the inner crust of the neutron star \cite{Chamel:2004in, Bulgac:2000qk},
which requires a quantum mechanical treatment with the band theory.
As shown in Refs. \cite{Watanabe:2017nzj, Minami:2022vch}, the Cooper pairs could influence the entrainment effect in the inner crust and alter the effective mass of free neutrons.
Hence, to describe the pasta phases in neutron stars, one needs to generalize the FOE method.

The purpose of the present paper is to develop the FOE method for the Hartree-Fock-Bogoliubov (HFB) Hamiltonian with a periodic potential.
We demonstrate validity and performance of the method for
a simple model of the slab phase in the inner crust of neutron stars.

The structure of this paper is as follows:
Sec.~\ref{ssec:FTHFB} reviews the finite-temperature HFB theory.
Section~\ref{ssec:densities} establishes an identity between the generalized density matrix and the Fermi-Dirac distribution function of the HFB Hamiltonian.
Section~\ref{ssec:band-theory} generalizes the identity to band theory.
Section~\ref{ssec:FOE} explains the FOE method to compute the generalized density matrix,
then applied to the 1D slab phase in Sec.~\ref{ssec:to-1D}.
In Sec.~\ref{sec:results}, 
the performance of the FOE method for the HFB band theory is studied with numerical calculations.
The nearsighted behavior of the numerical results is also discussed.
Finally, the conclusion is given in Sec.~\ref{sec:concl}.

\section{Fermi Operator Expansion for the HFB band theory}


\subsection{Finite-temperature HFB theory}
\label{ssec:FTHFB}

We recapitulate the HFB theory at finite temperature for many fermion systems.
Starting from an energy density functional $E[R]$,
where $R$ is the generalized density \cite{RS80,BR86}
including the normal density $\rho$ and the pair (abnormal) density $\kappa$,
the HFB Hamiltonian is given as
$(H_{\rm HFB})_{ij}\equiv\delta E'/ \delta R_{ji}$
with $E'[R]\equiv E[R]-\mu (\mathrm{tr}[\rho]-N_0)$.
The HFB equation in the coordinate-space representation is given by
\begin{equation}
\sum_{\sigma'}\int d\bm{r}'
	H_{\text{HFB}}(\bm{r}\sigma,\bm{r}'\sigma')
    \left(\begin{array}{c}
        u_{\nu}(\bm{r}'\sigma')\\
        v_{\nu}(\bm{r}'\sigma')
    \end{array}\right)=\epsilon_{\nu}\left(\begin{array}{c}
        u_{\nu}(\bm{r}\sigma)\\
        v_{\nu}(\bm{r}\sigma)
    \end{array}\right),
    \label{eq:HFBeq-general}
\end{equation}
where $\bm{r}$ and $\sigma=\pm 1/2$ represent the coordinate and the spin, respectively.
The HFB Hamiltonian in Eq.~(\ref{eq:HFBeq-general}) consists of 
the single-particle Hamiltonian $h$ and the pair potential $\Delta$
in a $2\times 2$ form as
\begin{equation}
	H_{\text{HFB}}\equiv
	\begin{pmatrix}
		h-\mu & \Delta \\
		-\Delta^* & -(h-\mu)^*
	\end{pmatrix}
	,
\end{equation}
where $\mu$ represents the chemical potential.
The HFB equations (\ref{eq:HFBeq-general}) define
the quasiparticle energy, $\epsilon_{\nu}$, and the quasiparticle wave functions,
$(u_{\nu}(\bm{r}\sigma),v_{\nu}(\bm{r}\sigma))$.
The quasiparticle creation and annihilation operators
$(\gamma_\nu,\gamma_\nu^\dagger)$ are given
by the following Bogoliubov transformation:
\begin{align}
& \psi(\bm{r}\sigma)=\sum_{\nu>0}\bigl[u_{\nu}(\bm{r}\sigma)\gamma_{\nu}+v_{\nu}^{*}(\bm{r}\sigma)\gamma_{\nu}^{\dagger}\bigr],
    \label{eq:Bogoliubov1} \\
& \psi^{\dagger}(\bm{r}\sigma)=\sum_{\nu>0}\bigl[u_{\nu}^{*}(\bm{r}\sigma)\gamma_{\nu}^{\dagger}+v_{\nu}(\bm{r}\sigma)\gamma_{\nu}\bigr],
    \label{eq:Bogoliubov2}
\end{align}
where $\psi^{\dagger}(\bm{r}\sigma)$ and $\psi(\bm{r}\sigma)$ are
the creation and annihilation field operators.
Here and hereafter,
the summation over the quasiparticles, $\sum_{\nu>0}$,
is restricted to those with positive energies, $\epsilon_\nu>0$.
The summation of all the quasiparticles
including those with negative energies will be
denoted as $\sum_{\nu\gtrless 0}$.
It is well known that all the solutions of Eq.~(\ref{eq:HFBeq-general})
with negative energies ($\nu<0$)
are expressed in terms of the positive energy solutions ($\nu>0$) as
\begin{equation}
	\begin{pmatrix}
		u_{-\nu}(\bm{r}\sigma) \\
		v_{-\nu}(\bm{r}\sigma)
	\end{pmatrix}
	=
	\begin{pmatrix}
		v^*_{\nu}(\bm{r}\sigma) \\
		u^*_{\nu}(\bm{r}\sigma)
	\end{pmatrix} , \quad\quad
	\epsilon_{-\nu}=-\epsilon_\nu \quad (\nu>0).
	\label{eq:negative-energy-solutions}
\end{equation}

\subsection{Densities in the finite-temperature HFB}
\label{ssec:densities}

At finite temperature, the normal and the pair densities are (see Refs. \cite{Kashiwaba:2020iiv} for derivation)
\begin{align}
    &\rho(\bm{r}\sigma,\bm{r}'\sigma') = 
	\langle \psi^{\dagger}(\bm{r}'\sigma') \psi(\bm{r}\sigma) \rangle_T\nl
    &\qquad =\sum_{\nu>0}\biggl[f(\epsilon_{\nu})u_{\nu}(\bm{r}\sigma)u_{\nu}^{*}(\bm{r}'\sigma')\nl
    &\qquad+\bigl(1-f(\epsilon_{\nu})\bigr)v_{\nu}^{*}(\bm{r}\sigma)v_{\nu}(\bm{r}'\sigma')\biggr], \label{eq:normal-den}\\
    &\kappa(\bm{r}\sigma,\bm{r}'\sigma') = \langle \psi(\bm{r}'\sigma') \psi(\bm{r}\sigma) \rangle_T\nl
    &\qquad =\sum_{\nu>0}\biggl[f(\epsilon_{\nu})u_{\nu}(\bm{r}\sigma)v_{\nu}^{*}(\bm{r}'\sigma')\nl
    &\qquad+\bigl(1-f(\epsilon_{\nu})\bigr)v_{\nu}^{*}(\bm{r}\sigma)u_{\nu}(\bm{r}'\sigma')\biggr],\label{eq:pair-den}
\end{align}
with $\langle\cdot\rangle_T$ as the thermal average and $f(x)=1/[1+\exp(x/T)]$ as the Fermi-Dirac distribution function.
The local densities are defined as
$\rho(\bm{r})=\sum_\sigma\rho(\bm{r}\sigma,\bm{r}\sigma)$
and
$\kappa(\bm{r})=\kappa(\bm{r}+1/2,\bm{r}-1/2)$.

Using the property of the HFB quasiparticle states,
Eq.~(\ref{eq:negative-energy-solutions}),
and that of the Fermi-Dirac function, $f(-x)=1-f(x)$,
the densities can be written as
\begin{align}
	\rho(\bm{r}\sigma,\bm{r}'\sigma') &= 
\sum_{\nu\gtrless 0} f(\epsilon_{\nu})u_{\nu}(\bm{r}\sigma)u_{\nu}^{*}(\bm{r}'\sigma') , \\
	\kappa(\bm{r}\sigma,\bm{r}'\sigma') &=
\sum_{\nu\gtrless 0}f(\epsilon_{\nu})u_{\nu}(\bm{r}\sigma)v_{\nu}^{*}(\bm{r}'\sigma') .
\end{align}
Thus, the generalized density matrix can be expressed as
\begin{align}
	R &=\begin{pmatrix}
		\rho & \kappa \\
		-\kappa^* & 1-\rho^*
	\end{pmatrix}
	=\sum_{\nu\gtrless 0} f(\epsilon_\nu)
	\begin{pmatrix}
		u_\nu \\
		v_\nu
	\end{pmatrix}
	\begin{pmatrix}
		u_\nu \\
		v_\nu
	\end{pmatrix}^\dagger \nl
	&=f(H_{\rm HFB})
	\sum_{\nu\gtrless 0}
	\begin{pmatrix}
		u_\nu \\
		v_\nu
	\end{pmatrix}
	\begin{pmatrix}
		u_\nu \\
		v_\nu
	\end{pmatrix}^\dagger
	 = f(H_{\rm HFB}) .
	\label{eq:generalized-density}
\end{align}
Note that we have the orthonormal and 
the completeness relations:
\begin{equation}
	\begin{pmatrix}
		u_\mu \\
		v_\mu
	\end{pmatrix}^\dagger
	\begin{pmatrix}
		u_\nu \\
		v_\nu
	\end{pmatrix}  = \delta_{\mu\nu} , \quad
	\sum_{\nu\gtrless 0}
	\begin{pmatrix}
		u_\nu \\
		v_\nu
	\end{pmatrix}
	\begin{pmatrix}
		u_\nu \\
		v_\nu
	\end{pmatrix}^\dagger  =
    1.
\label{eq:orthonormal-completeness}
\end{equation}
The important conclusion here is Eq.~(\ref{eq:generalized-density}), that
the generalized density matrix is nothing but
the Fermi-Dirac distribution function $f(x)$ with the argument $x$
replaced by $H_{\rm HFB}$.

\subsection{HFB band theory for generalized density}
\label{ssec:band-theory}

If the HFB Hamiltonian is invariant with respect to
a translation vector $\bm{T}$,
the index $\nu$ becomes a set of the Bloch vector $\bm{k}$
and the band index $n$.
The solution of Eq.~(\ref{eq:HFBeq-general})
can be written as
\begin{equation}
	\begin{pmatrix}
		u_{\nu} \\
		v_{\nu}
	\end{pmatrix}
	=
	e^{i\bm{k}\cdot\hat{\bm{x}}}
	\begin{pmatrix}
		\tilde{u}^{\bm{k}}_n \\
		\tilde{v}^{\bm{k}}_n
	\end{pmatrix} ,
	\label{eq:Bloch_wf}
\end{equation}
where $\hat{\bm{x}}$ is the coordinate operator.
Here, the wave functions $\tilde{u}^{\bm{k}}_n$ and $\tilde{v}^{\bm{k}}_n$ are
periodic for the translation $\bm{T}$,
namely,
$\tilde{u}^{\bm{k}}_n(\bm{r}+\bm{T},\sigma)=\tilde{u}_n(\bm{r}\sigma)$
and
$\tilde{v}^{\bm{k}}_n(\bm{r}+\bm{T},\sigma)=\tilde{v}_n(\bm{r}\sigma)$.
Using this property, we may reduce a problem in the large space
to that in the unit cell
with $N_k$ different Bloch $\bm{k}$
in the first Brillouin zone.
The HFB equation (\ref{eq:HFBeq-general}) can be cast as
\begin{equation}
	\sum_{\sigma'}\int d\bm{r}'
	H_{\text{HFB}}^{\bm{k}}(\bm{r}\sigma,\bm{r}'\sigma')
	\begin{pmatrix}
	    \tilde{u}^{\bm{k}}_n(\bm{r}'\sigma')\\
	    \tilde{v}^{\bm{k}}_n(\bm{r}'\sigma')
	\end{pmatrix}
	=\epsilon^{\bm{k}}_n\begin{pmatrix}
		\tilde{u}^{\bm{k}}_n(\bm{r}\sigma)\\
		\tilde{v}^{\bm{k}}_n(\bm{r}\sigma)
    \end{pmatrix} ,
    \label{eq:HFB-k}
\end{equation}
where $H_{\rm HFB}^{\bm{k}}$ is defined by the relation
$H_{\rm HFB}^{\bm{k}}=
e^{-i\bm{k}\cdot\hat{\bm{x}}} H_{\rm HFB}
e^{i\bm{k}\cdot\hat{\bm{x}}} $.
If all the potentials are local in the coordinate,
$H_{\rm HFB}^{\bm{k}}(\bm{r}\sigma)$ is identical to
$H_{\rm HFB}(\bm{r}\sigma)$ 
with the replacement of the derivatives as
$\nabla \rightarrow \nabla+i\bm{k}$.

The normalization condition,
the first relation of Eq.~(\ref{eq:orthonormal-completeness}),
is given in the coordinate space
\begin{equation}
\sum_\sigma \int_{V} d\bm{r} \left[
	\left| \tilde{u}^{\bm{k}}_n(\bm{r}\sigma) \right|^2
	+\left| \tilde{v}^{\bm{k}}_n(\bm{r}\sigma) \right|^2
		\right] = 1 ,
	\label{eq:normalization}
\end{equation}
where the integration is over the entire space $V=V_E$.
Alternatively, it is more convenient to
adopt the normalization in the unit cell,
namely, Eq.~(\ref{eq:normalization})
with $V$ being the volume of the unit cell.
Under this normalization,
the right hand side of Eq.~(\ref{eq:orthonormal-completeness})
should be multiplied by $N_k$.
Then, the generalized density of Eq. (\ref{eq:generalized-density})
in the coordinate space representation
should be modified to
\begin{align}
	R  = & \frac{1}{N_k}
	f(H_{\rm HFB})
	\sum_{\nu\gtrless 0}
	\begin{pmatrix}
		u^{\nu} \\
		v^{\nu}
	\end{pmatrix}
	\begin{pmatrix}
		u^{\nu} \\
		v^{\nu}
	\end{pmatrix}^\dagger  \nl
	& = \frac{1}{N_k}
	\sum_{\bm{k}}
	e^{i\bm{k}\cdot\hat{\bm{x}}}
	f(H_{\rm HFB}^{\bm{k}})
	\sum_{n\gtrless 0}
	\begin{pmatrix}
		\tilde{u}^{\bm{k}}_n \\
		\tilde{v}^{\bm{k}}_n
	\end{pmatrix}
	\begin{pmatrix}
		\tilde{u}^{\bm{k}}_n \\
		\tilde{v}^{\bm{k}}_n
	\end{pmatrix}^\dagger
	e^{-i\bm{k}\cdot\hat{\bm{x}}}
	.
	\label{eq:generalized-density-2}
\end{align}
For each $\bm{k}$, the operator
\begin{equation}
	\hat{P}_{\bm{T}} = 
	\sum_{n\gtrless 0}
	\begin{pmatrix}
		\tilde{u}^{\bm{k}}_n \\
		\tilde{v}^{\bm{k}}_n
	\end{pmatrix}
	\begin{pmatrix}
		\tilde{u}^{\bm{k}}_n \\
		\tilde{v}^{\bm{k}}_n
	\end{pmatrix}^\dagger ,
\end{equation}
gives the projection onto the subspace spanned
by periodic functions with respect to
the translation $\bm{T}$.
Since we easily find the relations,
$\left[ H_{\rm HFB}^{\bm{k}}, \hat{P}_{\bm{T}}\right]=0$
and $\hat{P}_{\bm{T}}^2=\hat{P}_{\bm{T}}$,
$f(H_{\rm HFB}^{\bm{k}})\hat{P}_{\bm{T}}$ is hermitian and
equal to
$\hat{P}_{\bm{T}}f(H_{\rm HFB}^{\bm{k}})\hat{P}_{\bm{T}}$.
Thus, Eq.~(\ref{eq:generalized-density-2}) is rewritten as
\begin{equation}
	R=\frac{1}{N_k}\sum_{\bm{k}} R^{\bm{k}}
	\label{eq:generalized-density-3}
\end{equation}
where
\begin{equation}
	R^{\bm{k}}
	= e^{i\bm{k}\cdot\hat{\bm{x}}}
	f(\tilde{H}_{\rm HFB}^{\bm{k}}) 
	e^{-i\bm{k}\cdot\hat{\bm{x}}} ,
	\label{eq:R^k}
\end{equation}
and
\begin{equation}
	\tilde{H}_{\rm HFB}^{\bm{k}} \equiv
	\hat{P}_{\bm{T}}
	H_{\rm HFB}^{\bm{k}}
	\hat{P}_{\bm{T}} .
\end{equation}
Because of the presence of the projector $\hat{P}_{\bm{T}}$ here,
we can assume the periodic property when we calculate
the operation of the Hamiltonian $\tilde{H}_{\rm HFB}^{\bm{k}}$
on an arbitrary state.
See Sec.~\ref{ssec:FOE}.

\subsection{FOE}
\label{ssec:FOE}

The most straightforward way of calculating
the generalized density $R$ is
to diagonalize ${H}_{\text{HFB}}^{\bm{k}}$
for solution of Eq.~(\ref{eq:HFB-k}),
and to construct $R$ with Eq.~(\ref{eq:generalized-density-2})
replacing $f(H^{\bm{k}}_{\rm HFB})$
by $f(\epsilon^{\bm{k}}_n)$.
However, the diagonalization costs the computation time of $O(N^3)$.
It is impractical for the 3D calculations.
The FOE provides an alternative way of constructing $R$
without the diagonalization of the Hamiltonian.

The FOE is based on a polynomial approximation
for a scaled Fermi-Dirac function:
\begin{equation}
    \bar{f}(x) = f(\epsilon_r x + \epsilon_c),   
    \label{eq:scaled-FD}
\end{equation}
where $\epsilon_c = (\epsilon_{\max} + \epsilon_{\min})/2$, $\epsilon_r = (\epsilon_{\max} - \epsilon_{\min})/2$. $\epsilon_{\max}$ and $\epsilon_{\min}$ are the largest and smallest possible quasiparticle energy eigenvalue in the adopted model space for the calculation.
Here, the dimensionless variable $x\in[-1,1]$ is introduced
to allow us to use the Chebyshev polynomials
which represent an orthonormal basis in $-1<x<1$ with the following definition definition of the inner product:
\begin{equation}
    \langle T_n(x), T_m(x) \rangle =
    \int_{-1}^{1} \frac{dx}{\sqrt{1-x^2}} \,
    T_n(x) T_m(x) .
\end{equation}
The Chebyshev-polynomial expansion is known to be numerically stable.

In the expansion of
\begin{equation}
\bar{f}(x) = \frac{a_0}{2} + \sum_{k=1}^{N_{\rm cheb}} a_k T_k(x),
\label{eq:f(x)}
\end{equation}
the coefficients $a_k$ are easily calculated as
\begin{equation}
    a_k = \frac{2}{\pi}
    \int_{-1}^{1} \frac{dx}{\sqrt{1-x^2}}\,
    T_k(x)\bar{f}(x).
\end{equation}
The required maximum degree of the polynomial $N_{\rm cheb}$
strongly depends on the temperature \cite{Wu:2002,Nakatsukasa:2022hpp}.
Introducing a scaled Hamiltonian
\begin{equation}
    \mathcal{H}_{\text{HFB}}^{\bm{k}} =
    \frac{
	    \tilde{H}_{\text{HFB}}^{\bm{k}} - \epsilon_c I
    }{
        \epsilon_r
    },
\end{equation}
where $I$ is the unit matrix in the model space,
the Chebyshev-polynomial expansion is applied to
$f(\tilde{H}_{\text{HFB}}^{\bm{k}})$ in Eq.~(\ref{eq:R^k}):
\begin{align}
f(\tilde{H}_{\text{HFB}}^{\bm{k}}) &=
    \bar{f}({\mathcal{H}}_{\text{HFB}}^{\bm{k}})\nl
    &=
    \frac{a_0}{2} I
	+ \sum_{k=1}^{N_{\rm cheb}}
        a_k
        T_k({\mathcal{H}}_{\text{HFB}}^{\bm{k}}).
    \label{eq:FOE}
\end{align}

The matrix elements of the generalized density $R_{ij}$
are calculated as follows.
Let $\ket{j}$ be the unit column vector with $\delta_{ij}$ as the $i$-th element in a given representation.
Then, $R_{ij} = (1/N_k) \sum_{\bm{k}} R^{\bm{k}}_{ij}$ with
\begin{align}
	R^{\bm{k}}_{ij} & = \Braket{i|
	e^{i\bm{k}\cdot\hat{\bm{x}}}
	\bar{f}(\mathcal{H}_{\rm HFB}^{\bm{k}}) 
	e^{-i\bm{k}\cdot\hat{\bm{x}}} |j}  \nl
	&= \frac{a_0}{2} \delta_{ij} + \sum_{n=1}^{N_{\rm cheb}}
	a_n \Braket{i_0;\bm{k}|j_n;\bm{k}} ,
    \label{eq:Rij}
\end{align}
where $\ket{j_n;\bm{k}}\equiv T_n(
\mathcal{H}_{\text{HFB}}^{\bm{k}})
e^{-i\bm{k}\cdot\hat{\bm{x}}} \ket{j}$ with $n\geq 0$
are calculated using a recursion relation for
the Chebyshev polynomials as \cite{Nakatsukasa:2022hpp}
\begin{align}
& \ket{j_0;\bm{k}}=e^{-i\bm{k}\cdot\hat{\bm{x}}}\ket{j},
\quad\quad
\ket{j_1;\bm{k}}=\mathcal{H}_{\text{HFB}}^{\bm{k}}\ket{j_0;\bm{k}} , \label{eq:j0j1}\\
&\ket{j_n;\bm{k}}=2\mathcal{H}_{\text{HFB}}^{\bm{k}}\ket{j_{n-1};\bm{k}} 
	-\ket{j_{n-2};\bm{k}} .
    \label{eq:jn}
\end{align}
The normal and pair densities, $\rho$ and $\kappa$,
are obtained by
$\rho_{ij}=R_{ij}$ and
$\kappa_{ij}=-R_{N+i,j}^*$ with $1 \leq i,j \leq N$
where $N$ is the dimension of the single-particle space.

Based on Eqs.~\eqref{eq:j0j1} and \eqref{eq:jn}, one can independently perform the computation of each ket state specified by $j$ and $\bm{k}$.
The required number of $N_{\exp}$ does not depend on the space dimension. 
Therefore, the FOE is of time complexity $O(N^2)$, faster than directly solving the HFB equation, which is of $O(N^3)$.
This formulation allows one to effectively parallelize the computation by introducing independent threads or processes for each $j$.

Furthermore, one can further accelerate the computation if the densities are nearsighted.
The nearsightedness is usually defined by the localization of the density matrix $\rho(\bm{r},\bm{r}')$=0 at $|\bm{r}-\bm{r}'|>r_N$,
where $r_N$ is a characteristic nearsighted distance.
The nearsightedness is also confirmed in the nuclear matter
at finite temperature \cite{Nakatsukasa:2022hpp}.
If $\rho(\bm{r},\bm{r}')$ and $\kappa(\bm{r},\bm{r}')$ are {\it nearsighted}, when computing $\ket{j_n,\bm{k}}$,
one can calculate it in a space the dimension of which is smaller than $N$.
In other words, one can assume that all the components
$\braket{i|j_n}$ vanish for $|z_i-z_j|>r_N$.
Then, for the coordinate-space calculations,
the complexity of computing $\ket{j_n,\bm{k}}$ with given $j$ and $\bm{k}$
does not depend on the space dimension $N$, and
the time complexity becomes $O(N)$.
For a detailed discussion of the nearsightedness, see Sec.~\ref{ssec:nearsightedness}.

\begin{figure}
    \centering
    \includegraphics[width=0.6\linewidth]{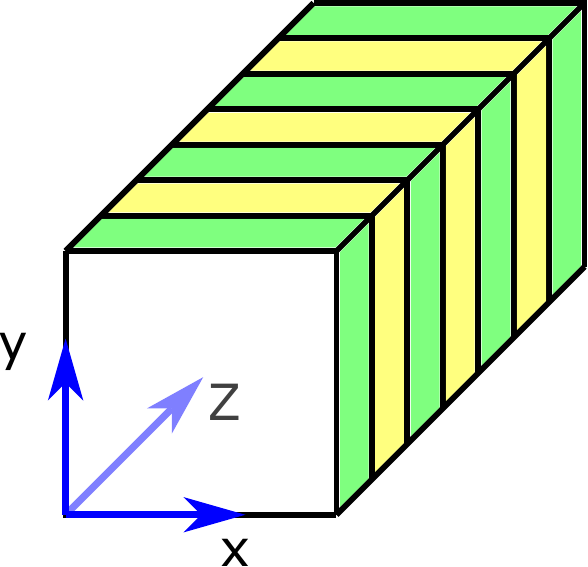}
    \caption{Schematic image of the slab phase.}
    \label{fig:schematic}
\end{figure}

\subsection{One-dimensional slab phase} 
\label{ssec:to-1D}

In this paper, we demonstrate the numerical results of the slab phase,
in which the nonuniform structure exists only for the $z$ direction
(see Fig.~\ref{fig:schematic}).
The wave functions are trivially given by the plane waves for
the $x$ and $y$ directions.
Thus, the wave functions in Eq.~(\ref{eq:Bloch_wf})
in the coordinate-space representation are
\begin{equation}
	\begin{pmatrix}
		u_n^{\bm{k}}(\bm{r}\sigma) \\
		v_n^{\bm{k}}(\bm{r}\sigma)
	\end{pmatrix}
	=
	e^{i\bm{k}\cdot{\bm{r}}}
	\begin{pmatrix}
		\tilde{u}^{\bm{k}}_n(z\sigma) \\
		\tilde{v}^{\bm{k}}_n(z\sigma)
	\end{pmatrix} .
\end{equation}
The normalization is introduced for the $z$ direction with a unit cell
of the length $L$.
\begin{equation}
    \sum_\sigma \int_{-L/2}^{L/2} dz \left( 
	|\tilde{u}_n^{\bm{k}}(z\sigma)|^2 + |\tilde{v}_n^{\bm{k}}(z\sigma)|^2
    \right) = 1.
\end{equation}
The HFB equation is given by the same as Eq.~(\ref{eq:HFB-k}).
If the potential in the single-particle Hamiltonian $h$ and
the pair potential $\Delta$ are local in the coordinate,
we may write them in the form
\begin{equation}
\sum_{\sigma'} H_{\text{HFB}}^{\bm{k}}(z;\sigma,\sigma')
	\begin{pmatrix}
        \tilde{u}_n^{\bm{k}}(z\sigma')\\
        \tilde{v}_n^{\bm{k}}(z\sigma')
	\end{pmatrix}
	=\epsilon_n^{\bm{k}}
	\begin{pmatrix}
        \tilde{u}_n^{\bm{k}}(z\sigma)\\
        \tilde{v}_n^{\bm{k}}(z\sigma)
	\end{pmatrix} ,
    \label{eq:HFBeq-bloch}
\end{equation}
with
\begin{equation}
H_{\text{HFB}}^{\bm{k}}(z;\sigma,\sigma') = 
	\begin{pmatrix}
h_{\sigma\sigma'}^{+\bm{k}}(z)-\mu & \Delta^{+\bm{k}}_{\sigma\sigma'}(z)\\
-\Delta_{\sigma\sigma'}^{-\bm{k}*}(z) & -h^{-\bm{k}*}_{\sigma\sigma'}(z)+\mu
    \end{pmatrix}
	.
    \label{eq:H-HFB-bloch}
\end{equation}
Here, $h_{\sigma\sigma'}^{\pm\bm{k}}(z)$ and
$\Delta^{\pm\bm{k}}_{\sigma\sigma'}(z)$
are given by
replacing 
differentiation $\partial_z$ ($\partial_{x,y}$)
in $h_{\sigma\sigma'}(\bm{r})$ and $\Delta_{\sigma\sigma'}(\bm{r})$
by $\partial_z\pm ik_z$ ($\pm ik_{x,y}$),
respectively.
If the derivatives are present only in the kinetic term in
$h_{\sigma\sigma'}$ as
$h_{\sigma\sigma'}=-\nabla^2/2m^*(z) + U_{\sigma\sigma'}(z)$,
$h_{\sigma\sigma'}^{\pm\bm{k}}(z)=-(\partial_z\pm ik_z)^2/(2m^*(z))
+k_\xi^2/(2m^*(z)) + U_{\sigma\sigma'}(z)$
and
$\Delta^{\bm{k}}_{\sigma\sigma'}(z)=\Delta_{\sigma\sigma'}(z)$,
where $m^*(z)$ is the effective mass and $k_\xi\equiv k_x^2+k_y^2$.
Note that the HFB Hamiltonian does not depend on
the direction of $\bm{k}$ in the $x$-$y$ plane.

Using the FOE formula of Eq.~(\ref{eq:generalized-density-3}),
the generalized density in the coordinate space is given by
\begin{align}
R(\bm{r}\sigma,\bm{r}'\sigma')
&=\frac{1}{N_k}\sum_{\bm{k}} R^{\bm{k}}(\bm{r}\sigma,\bm{r}'\sigma') \nl
	&=\frac{1}{N_k}
	\sum_{\bm{k}} e^{ik_x(x-x')+ik_y(y-y')} \nl
	&\quad\quad \times \left[
	e^{ik_z z} \bar{f}\left(\mathcal{H}_{\rm HFB}^{\bm{k}} \right)
	e^{-ik_z z'} \right]_{z\sigma,z'\sigma'} \nl
&=\frac{1}{N_{k_\xi}N_{k_z}} \sum_{k_\xi}J_0(k_\xi \xi) \nl
	&\quad \quad \times \sum_{k_z} \left[
	e^{ik_z z} \bar{f}\left(\mathcal{H}_{\rm HFB}^{\bm{k}} \right)
	e^{-ik_z z'} \right]_{z\sigma,z'\sigma'} ,
    \label{eq:R-final}
\end{align}
where $N_k=N_{k_x} N_{k_y} N_{k_z}=N_{k_\xi} N_{k_\theta} N_{k_z}$
and $\xi\equiv\sqrt{(x-x')^2+(y-y')^2}$.
Here, we use the fact that $\mathcal{H}_{\rm HFB}^{\bm{k}}$
does not depend on $k_\theta$, which allows us to
integrate over the orientation of $\bm{k}$ in the $x$-$y$ plane.
From the above expression, it is straightforward to see that in the one-dimensional slab phase, the local densities,
$\rho(\bm{r})=\sum_\sigma \rho(\bm{r},\sigma;\bm{r},\sigma)$
and
$\kappa(\bm{r})=\kappa(\bm{r},+1/2;\bm{r},-1/2)
=-\kappa(\bm{r},-1/2;\bm{r},+1/2)$,
do not depend on $(x,y)$.
We denote them as $\rho(z)$ and $\kappa(z)$.

\section{Numerical method and results}
\label{sec:results}

To study the usefulness of the FOE for the HFB band theory,
we adopt a simple model of the slab phase and
perform numerical calculations.
The single-particle potential is given in the Woods-Saxon form,
and the zero-range interaction of the delta function form
is assumed for the pairing interaction.

\subsection{Setup and algorithm}
\label{ssec:setup}

For most calculations in this paper,
the temperature is set to $T=100$ keV.
Realistic temperatures for neutron stars are estimated as
$T=0.01$--$1\unit{MeV}$,
depending on the age of the neutron stars.
We are interested in the inner crust with the free neutrons,
therefore, we choose the chemical potential to be positive, $\mu=10\unit{MeV}$ in the following studies.
The 1D spin-independent potential acting on the neutrons are
\begin{equation}
    U_{\sigma\sigma'}(z) = -\frac{\delta_{\sigma\sigma'} U_0}{1+e^{(|z|-z_0)/a}},\quad -L/2<z<L/2,
\end{equation}
with periodicity
\begin{equation}
    U_{\sigma\sigma'}(z) = U_{\sigma\sigma'}(z+L).
\end{equation}
Here, $U_0=50.00\unit{MeV}$, $z_0=10.00\unit{fm}$, $a=2.00\unit{fm}$, and $L=48.00\unit{fm}$.
We choose the mesh size of the adopted model space as $\Delta z = 0.4 \unit{fm}$.
Due to the finite mesh size, the square of the maximum momentum
of the neutron is approximately $3(\pi\hbar/\Delta z)^2$, so we may assume $\epsilon_{\max} \approx 3\pi^2\hbar^2/(2m \Delta z^2)$ and $\epsilon_{\min} \approx -3\pi^2\hbar^2/(2m \Delta z^2)$, which leads to $\epsilon_r=\epsilon_{\max}$ and $\epsilon_c=0$ in Eq. (20).
We find that we can further reduce the value of $\epsilon_r$ without losing precision, and in this study we adopt $\epsilon_r=2.5\times \hbar^2/(m \Delta z^2) \approx 647.98 \unit{MeV}$.

We calculate the pair potential $\Delta_{\sigma\sigma'}(z)$ self-consistently, and require it to satisfy
\begin{equation}
    \Delta_{\sigma\sigma'}(z)=2\sigma\Delta(z)\delta_{\sigma,-\sigma'},
    \label{eq:Dij}
\end{equation}
where $\Delta(z)$ is a scalar function that depends on the local pair density as
\begin{equation}
    \Delta(z) = \int_{-L/2}^{L/2} g \kappa(z') \delta(z-z').
    \label{eq:D-g-kappa}
\end{equation}
Here, we choose $g=150$--$200\unit{MeV\cdot fm^3}$ to ensure the average pair gap to be
\begin{equation}
    |\bar{\Delta}|= \frac{\int_{-L/2}^{L/2}dz\,\rho(z) |\Delta(z)|}{\int_{-L/2}^{L/2}dz\,\rho(z)} \approx 1.00\unit{MeV},
\end{equation}
which is a realistic value of the pair gap that could occur in the inner crust of a neutron star.

The algorithm for the FOE to obtain $\rho(\bm{r}\sigma,\bm{r}'\sigma')$
and $\kappa(\bm{r}\sigma,\bm{r}\sigma')$ is as follows.
\begin{enumerate}
    \item Start with some initial guess of $\Delta(z)$, and construct $\mathcal{H}_{\text{HFB}}^{\bm{k}}$ satisfying Eq.~\eqref{eq:H-HFB-bloch}.
    \label{step:solve-H}
    \item Use the FOE, Eq.~\eqref{eq:Rij}, to calculate $R^{\bm{k}}_{ij}$ from $\mathcal{H}_{\text{HFB}}^{\bm{k}}$.
    \label{step:expansion}
    \item Extract $\rho(\bm{r}\sigma,\bm{r}\sigma')$ and $\kappa(\bm{r}\sigma,\bm{r}\sigma')$ from $R^{\bm{k}}_{ij}$.
    \label{step:solve-rho-kappa}
    \item Update the $\Delta(z)$ in $\mathcal{H}_{\text{HFB}}^{\bm{k}}$ using the new local pair density following Eq.~\eqref{eq:D-g-kappa}.
    Use $\Delta(z) = (1-\alpha) \Delta^{\text{(old)}}(z) + \alpha\Delta^{\text{(new)}}(z)$ with $\alpha=0.5$.
    \label{step:update-H}
    \item Repeat procedures \ref{step:expansion}--\ref{step:update-H} iteratively, until the normal and pair densities converge.
\end{enumerate}

In this algorithm, we need to determine $N_{\rm cheb}$ in Eq.~\eqref{eq:Rij}.
For this purpose, we define $\rho_j^{(n)}$ and $\kappa_j^{(n)}$ as the local normal and pair densities computed with $n$ terms of Chebyshev expansion at $z=z_j$, and introduce
\begin{eqnarray}
    \delta\rho_{j}^{(N_{\text{cheb}})}&=&\rho_{j}^{(N_{\text{cheb}})}-\rho_{j}^{(N_{\text{cheb}}-n_{0})}
    \label{eq:delta_rho} \\
    \delta\kappa_{j}^{(N_{\text{cheb}})}&=&\kappa_{j}^{(N_{\text{cheb}})}-\kappa_{j}^{(N_{\text{cheb}}-n_{0})}.
    \label{eq:delta_kappa}
\end{eqnarray}
Here, $n_0$ is a positive integer smaller than $N_{\text{cheb}}$.
During the calculation of the FOE, on each space lattice site $j$, we evaluate $\delta\rho_{j}^{(N_{\text{cheb}})}$ and $\delta\kappa_{j}^{(N_{\text{cheb}})}$
increasing the maximum degree of the Chebyshev polynomial $N_\text{cheb}$ by $n_0$.
Once $\delta\rho_{j}^{(N_{\text{cheb}})}<10^{-5}\unit{fm^{-3}}$ and $|\delta\kappa_{j}^{(N_{\text{cheb}})}|<10^{-5}\unit{fm^{-3}}$, we terminate the expansion on the site.
This method allows us to use different values of $N_{\text{cheb}}$ for different sites.
To save computation time, one also needs to choose $n_0\gg 1$.
In the present calculation, we use $n_0=100$.


\subsection{Performance of the FOE for the HFB band theory}
\label{ssec:validity}

In this subsection, we compare the numerical results of the FOE method
with those of the diagonalization method.
Here, the diagonalization method means to replace 
steps~\ref{step:expansion} and \ref{step:solve-rho-kappa} in the previous subsection with solving Eq.~\eqref{eq:HFBeq-bloch} by diagonalization and computing $\rho$ and $\kappa$ using Eqs.~\eqref{eq:normal-den} and \eqref{eq:pair-den}.

\begin{figure}
    \centering
    \includegraphics[width=\linewidth]{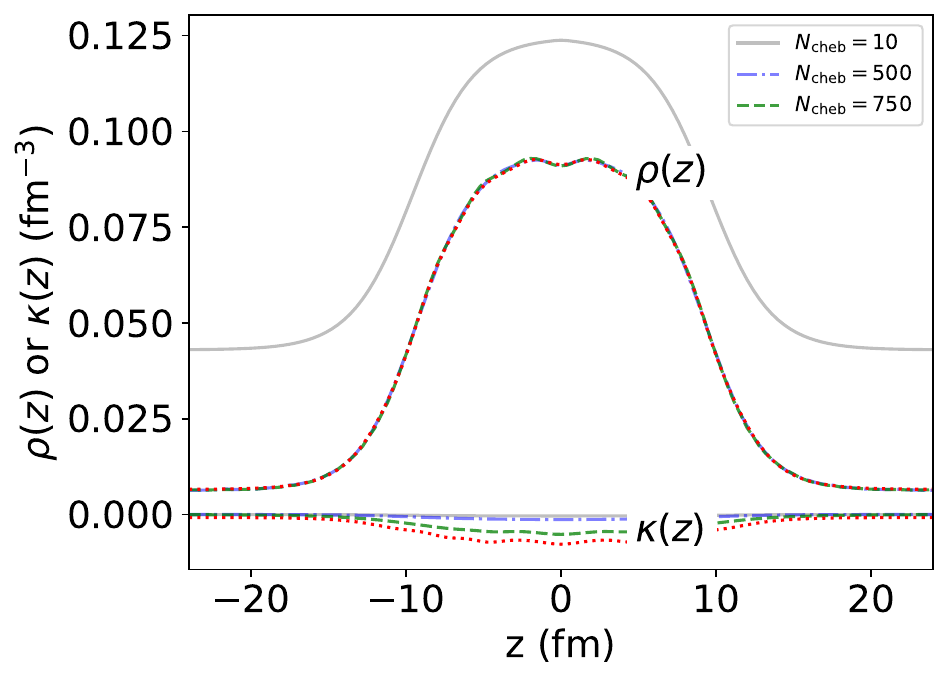}
    \caption{Calculated normal densities $\rho(z)$ and pair densities $\kappa(z)$ at $T=0.1$ MeV
    using the FOE method with fixed $N_{\text{cheb}}$.
    The red dotted line is the result of the matrix
    diagonalization method.
    The average pairing gap is $|\bar{\Delta}|=0.89\unit{MeV}$.
}
    \label{fig:N-rho-kappa}
\end{figure}

\begin{figure}
    \centering
    \includegraphics[width=\linewidth]{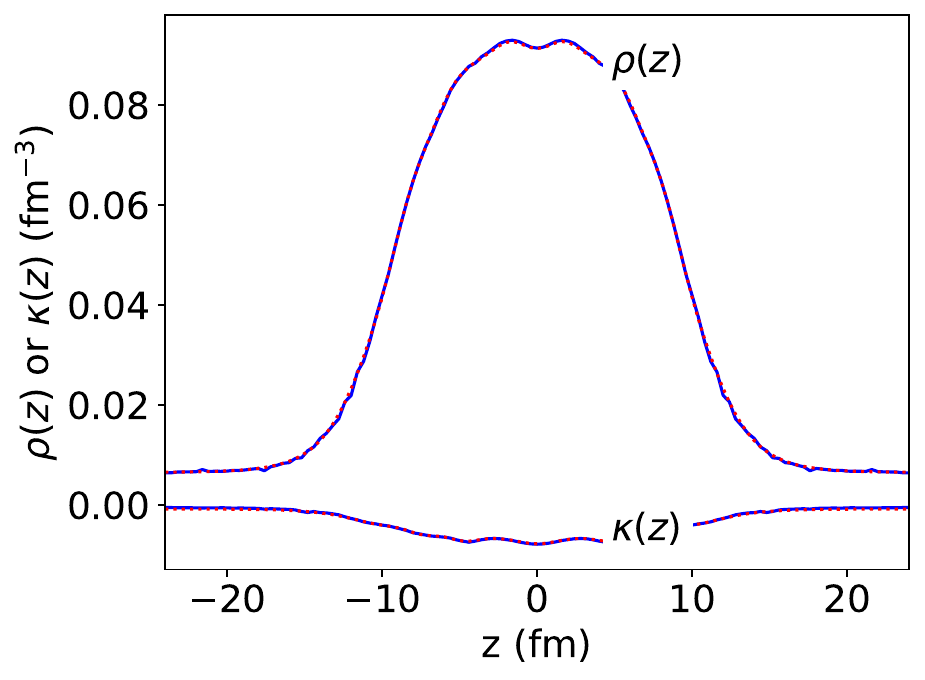}
    \caption{The same as Fig.~\ref{fig:N-rho-kappa} but
    adopting $N_\text{cheb}$ satisfying $\delta\rho_{j}^{(N_{\text{cheb}})}<10^{-5}\unit{fm^{-3}}$ and $|\delta\kappa_{j}^{(N_{\text{cheb}})}|<10^{-5}\unit{fm^{-3}}$ as described in Sec.~\ref{ssec:setup}.
}
    \label{fig:FOE_HFB}
\end{figure}

We plot the densities, $\rho(z)$ and $\kappa(z)$, in Figs.~\ref{fig:N-rho-kappa} and \ref{fig:FOE_HFB}.
In Fig.~\ref{fig:N-rho-kappa}, the FOE results with fixed values of $N_{\text{cheb}}=10, 500$, and 750 are shown.
With an increase of $N_{\text{cheb}}$, both $\rho(z)$ and $\kappa(z)$ steadily approach the results of the diagonalization method.
Below a threshold value of $N_{\text{cheb}}$, $\kappa(z)$ vanishes.
This is because when $N_{\text{cheb}}$ is small,
the function of Eq.~(\ref{eq:f(x)})
is flatter than the accurate Fermi-Dirac distribution,
which corresponds to a higher effective temperature.
The normal density $\rho(z)$ quickly reaches
the convergence, while the pair density $\kappa(z)$ requires larger values
of $N_\text{cheb}$.

In Fig.~\ref{fig:FOE_HFB}, we determine $N_{\text{cheb}}$ following the condition of $\delta\rho_{j}^{(N_{\text{cheb}})}<10^{-5}\unit{fm^{-3}}$ and $|\delta\kappa_{j}^{(N_{\text{cheb}})}|<10^{-5}\unit{fm^{-3}}$,
described in Sec.~\ref{ssec:setup}.
$N_{\text{cheb}}$ depends on the coordinate $z$, and
we have $N_{\text{cheb}} \simeq 2500$ near $z=0$.
A similar number of $N_\text{cheb}$ is reported in
the KS theory without pairing~\cite{Nakatsukasa:2022hpp}.

Next we investigate performance of the FOE method with different temperatures and pairing strengths.
We define the average variation of pair potential with increasing
$N_\text{cheb}$ as
\begin{equation}
   \left|\delta\bar{\Delta}^{(N_{\text{cheb}})}\right|=\left|\frac{\sum_{j}g\delta\kappa_{j}^{(N_{\text{cheb}})}\rho_{j}}{\sum_{j}\rho_{j}}\right|,
    \label{eq:average_error}
\end{equation}
where the summation is taken over all the lattice sites $j$,
and $\rho_j$ is the local normal density on that site.
$\delta\kappa_j^{(N_\text{cheb})}$
is given by Eq.~\eqref{eq:delta_kappa}
with $n_0=100$.

\begin{figure}
    \centering
    \includegraphics[width=\linewidth]{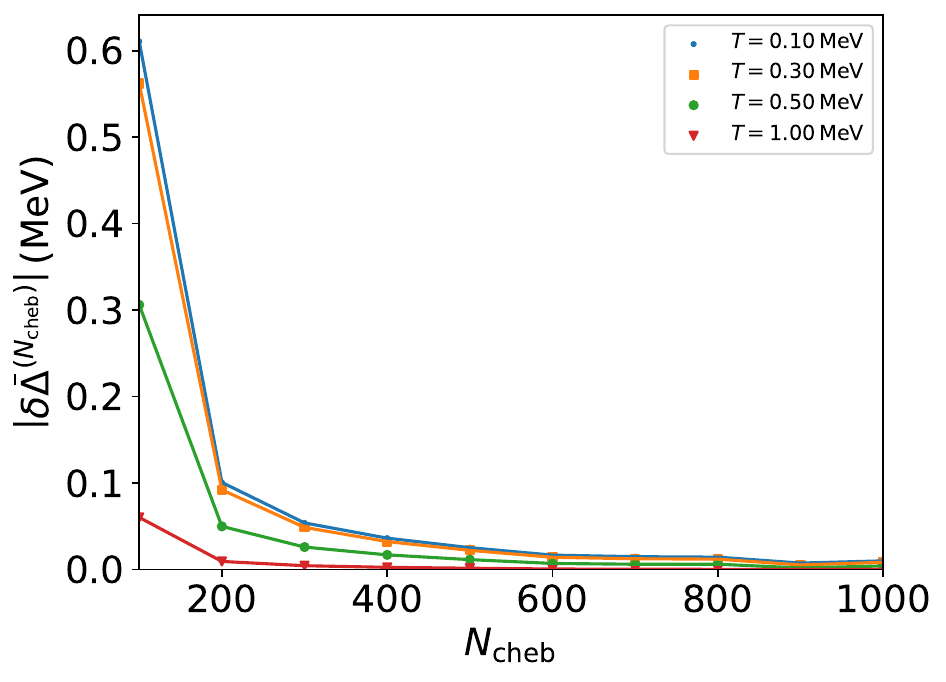}
    \caption{The average variation of pair potential, $|\delta\bar{\Delta}^{(N_{\text{cheb}})}|$
    in \eqref{eq:average_error},
    as a function of $N_{\text{cheb}}$,
    under different temperatures.
    We fix the pairing coupling strength as $g=155.51\unit{MeV\cdot fm^3}$.}
    \label{fig:T-err}
\end{figure}

In Fig.~\ref{fig:T-err}, the average variation of the pair potential as a function of
$N_\text{cheb}$ is shown.
We fix the pairing strength $g=155.51\unit{MeV\cdot fm^3}$.
The calculated average pair gaps are
$|\bar{\Delta}|=0.89, 0.85, 0.56, 0.09\unit{MeV}$ for $T=0.10, 0.30, 0.50$, and $1.00\unit{MeV}$, respectively.
At even lower temperature of $T=0$ and $0.01\unit{MeV}$,
we find the result almost identical to the one
at $T=0.10\unit{MeV}$, so we do not show them
in Fig.~\ref{fig:T-err}.
These results show that $|\delta\bar{\Delta}^{(N_{\text{cheb}})}|$ decreases much faster with $N_{\text{cheb}}$
at higher temperature.
Hence, we need larger values of $N_\text{cheb}$ at lower temperatures.
This is because the Fermi-Dirac distribution function
at low temperature is close to the step function.
Thus, one needs the larger maximum degree of Chebyshev polynomials
to approximate the distribution.
However, we have verified that the required values of $N_{\rm cheb}$ do not diverge at $T=0$.
This is natural since the finite pair potential produces a gap in the quasiparticle energy spectra near $\epsilon=0$ where the Fermi-Dirac distribution $f(\epsilon)$ with $T=0$ has a discontinuity.


\begin{figure}
    \centering
    \includegraphics[width=\linewidth]{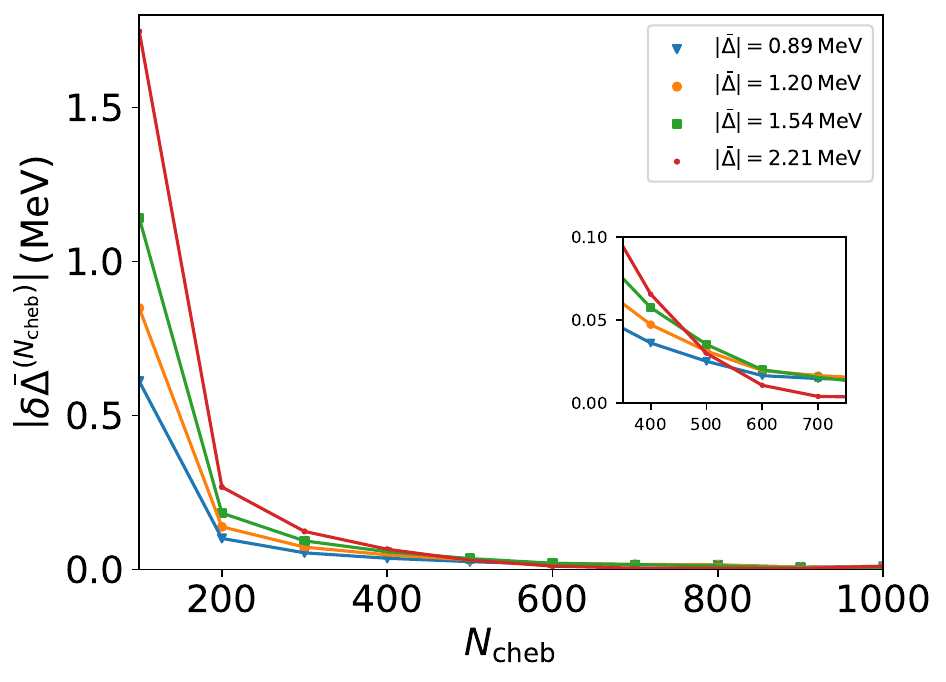}
    \caption{The same as Fig.~\ref{fig:T-err}, but
    with fixed temperature 
    $T=0.10\unit{MeV}$ and different pairing force strengths $g$.
    }
    \label{fig:Delta-err}
\end{figure}

\begin{figure*}
    \centering
    \begin{subfigure}{0.49\linewidth}
        \includegraphics[width=\linewidth]{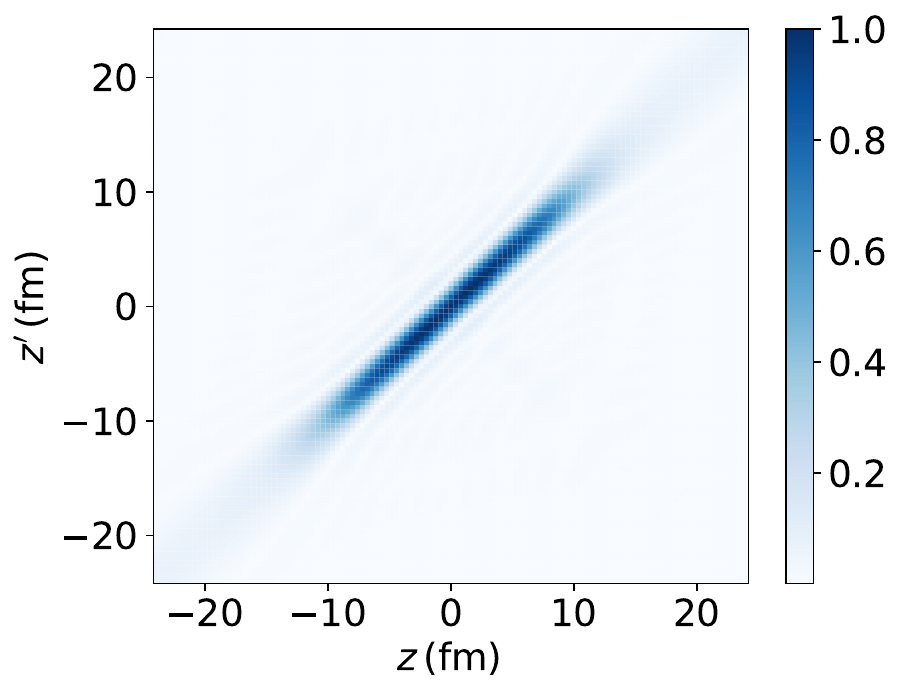}
        \subcaption{}
    \end{subfigure}
    \begin{subfigure}{0.49\linewidth}
        \includegraphics[width=\linewidth]{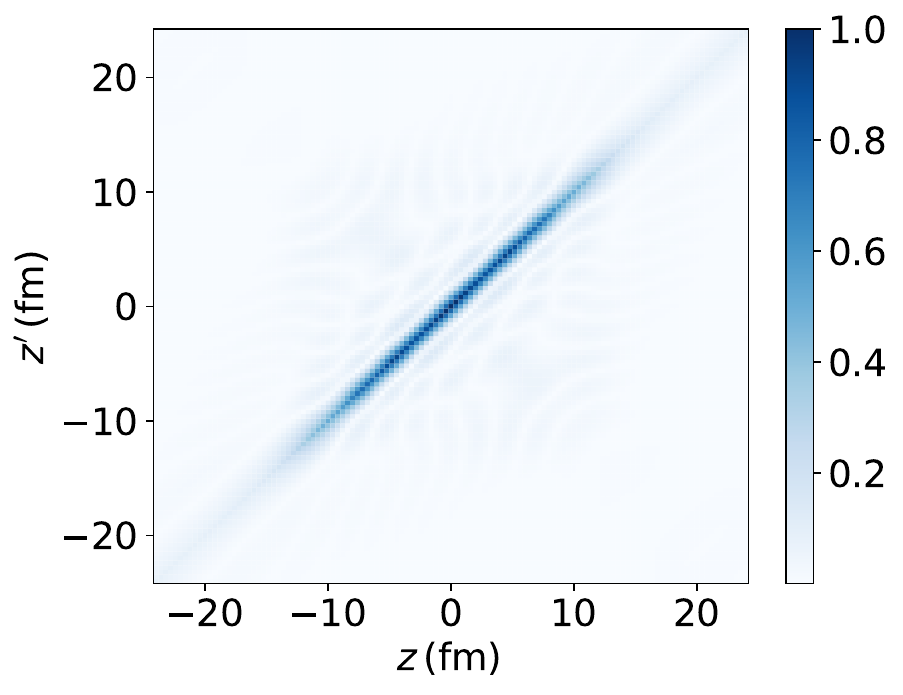}
        \subcaption{}
    \end{subfigure}
    \caption{Calculated density matrices (a) $\rho(z,z')$ and (b) $\kappa(z,z')$ with $x=x'$ and $y=y'$,
    at $T=0.1\unit{MeV}$ and $|\bar{\Delta}|=0.89\unit{MeV}$.
    The maximum values of the distributions are normalized to unity.}
    \label{fig:rho-kappa-ns-z}
\end{figure*}

\begin{figure}
    \centering
    \includegraphics[width=\linewidth]{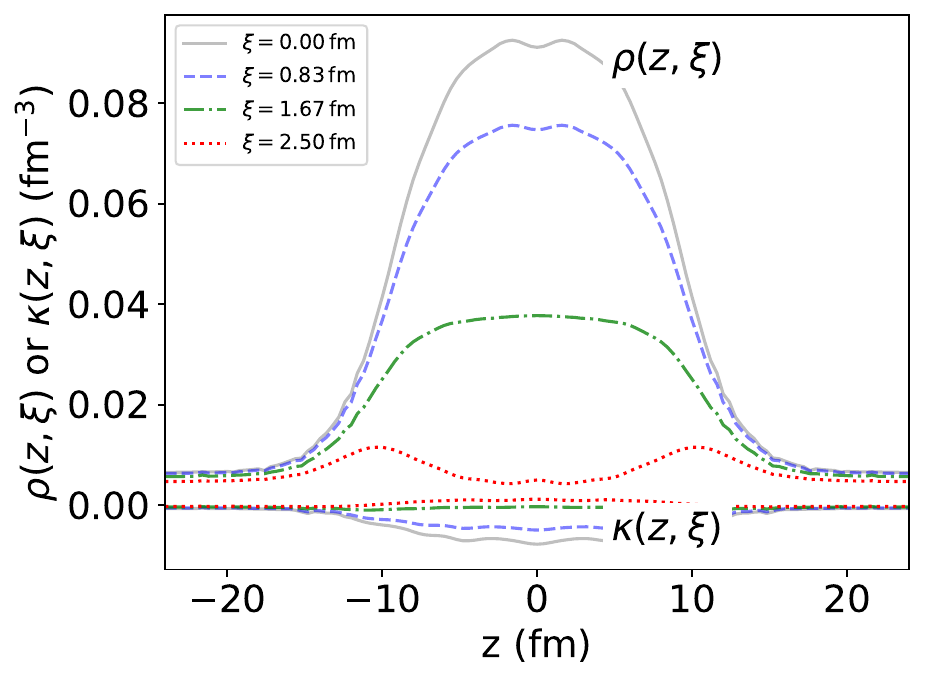}
    \caption{Calculated density matrices, $\rho(z,\xi)$ and $\kappa(z,\xi)$ with $z=z'$,
    at $T=0.1\unit{MeV}$ and $|\bar{\Delta}|=0.89\unit{MeV}$,
    for different values of $\xi=\sqrt{(x-x')^2+(y-y')^2}$.
    }
    \label{fig:rho_kappa_xi}
\end{figure}

\begin{figure}
    \centering
    \includegraphics[width=\linewidth]{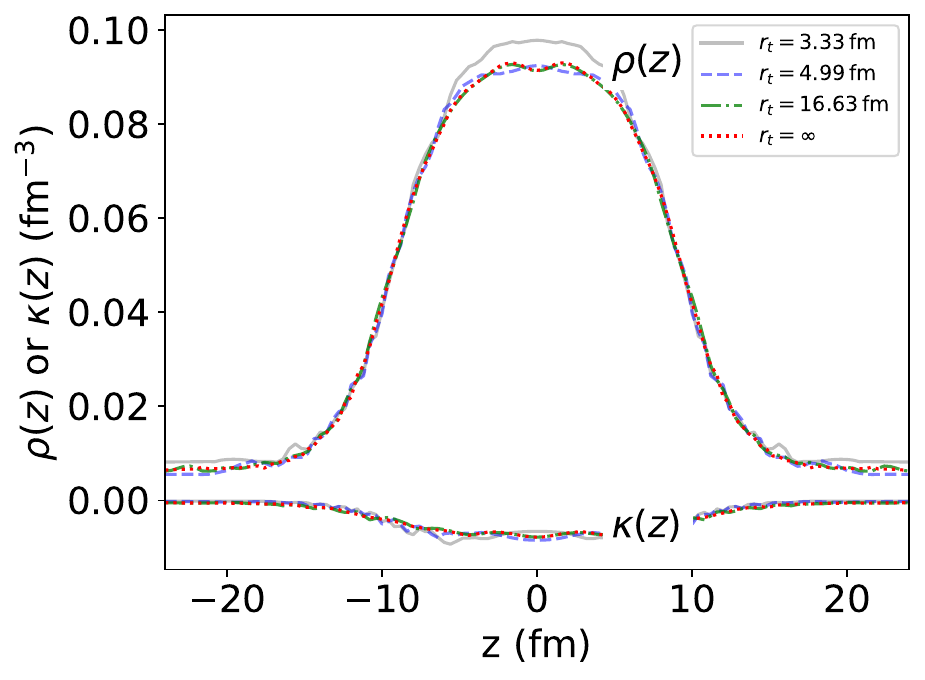}
    \caption{Normal density $\rho(z)$ and pair density $\kappa(z)$ 
    at $T=0.1\unit{MeV}$ and $|\bar{\Delta}|=0.89\unit{MeV}$
    with different truncation distance $r_t$.
}
    \label{fig:rho_kappa_rn}
\end{figure}

We fix the temperature $T=0.1\unit{MeV}$,
and vary the pairing strength $g$,
which produces different average pairing gaps,
$|\bar{\Delta}|=0.89, 1.20, 1.54$, and $2.21\unit{MeV}$.
The variations of pair potential with respect to $N_{\text{cheb}}$
are shown in Fig.~\ref{fig:Delta-err}.
We find that,
at small values of $N_{\text{cheb}}$,
$|\delta\bar{\Delta}^{(N_{\text{cheb}})}|$
is larger for larger values of the average pairing gap.
However, at large values of $N_{\text{cheb}}$,
the ordering of $|\delta\bar{\Delta}^{(N_{\text{cheb}})}|$
becomes opposite,
smaller $|\delta\bar{\Delta}^{(N_{\text{cheb}})}|$
for larger $|\bar{\Delta}|$ (as is shown on the inset).
Therefore, increasing the pair gap leads to a smaller error
in the Chebyshev expansion.
This is expected as the pair potential measures
the gap between the positive and negative quasiparticle energies.
Under the presence of larger gap,
the required $N_\text{cheb}$ could be smaller 
because an inaccurate description in the gap region
does not increase the error.

In conclusion, for the inner crust of neutron stars,
the FOE is a useful method to solve the HFB band theory.
Especially, it is more favorable for
higher temperature and stronger pairing.

\subsection{Nearsightedness}
\label{ssec:nearsightedness}

In the closing part of Sec.~\ref{ssec:FOE}, we show that one can accelerate the calculation if the densities are nearsighted.
In this subsection, we show that $\rho(\bm{r},\bm{r}')$ and $\kappa(\bm{r},\bm{r}')$ are in fact {\it nearsighted}.

We denote $\rho(z,z') = \rho(\bm{r},\bm{r}')|_{x=x',y=y'}$, $\kappa(z,z') = \kappa(\bm{r},\bm{r}')|_{x=x',y=y'}$.
The nearsightedness in the $z$ direction is shown
in Fig.~\ref{fig:rho-kappa-ns-z},
with two-dimensional graphs of $\rho(z,z')$ and $\kappa(z,z')$,
calculated at $T=0.10\unit{MeV}$.
Both the normal and pair densities are nearsighted.
The characteristic length of the nearsightedness $r_N$
is several femtometers.
This value of $r_N$ is qualitatively in agreement with
the value for the KS theory 
\cite{Baer1997, Nakatsukasa:2022hpp};
$r_N\approx\sqrt{\hbar^2/(3mT)}\approx 10\unit{fm}$
for $T=0.1\unit{MeV}$.
We also examine the temperature dependence of $r_N$
and confirm the behavior as $r_N\approx\sqrt{\hbar^2/(3mT)}$. 
The present formula for $r_N$ at finite temperature
is qualitatively valid for the HFB band theory.
In Fig.~\ref{fig:rho-kappa-ns-z}, it is visible that
the damping of the off-diagonal elements is stronger
in the central region $|z|\lesssim 10$ fm
than in the outer region $|z|\gtrsim 10$ fm.
This suggests that the nearsightedness is stronger
at higher density.

The nearsightedness in the transverse ($x$ and $y$) directions
is shown in Fig.~\ref{fig:rho_kappa_xi}.
From Eq.~\eqref{eq:R-final},
instead of the four coordinates $x,y$ and $x',y'$,
the densities depend only on $\xi=\sqrt{(x-x')^2+(y-y')^2}$.
To investigate the transverse nearsightedness,
we define the quantities
$\rho(z,\xi)\equiv\rho(\bm{r},\bm{r}')|_{z=z'}$ and
$\kappa(z,\xi)\equiv\kappa(\bm{r},\bm{r}')|_{z=z'}$.
In the central region ($|z|\lesssim 10$ fm),
the magnitude of density values quickly vanishes,
$\rho,\kappa\rightarrow 0$,
as the transverse off-diagonal distance $\xi$ increases.
In contrast, in the outer low-density region ($|z|\gtrsim 10$ fm),
the damping behavior with respect to $\xi$ is weaker.
This is analogous to the density dependence of the nearsightedness
in the $z$ direction.
The nearsighted length $r_N$ is of a few femtometers,
which is similar to that of the $z$ direction.
Hence, the nearsighted features are approximately common
in the $z$ and transverse directions.

Finally, we use the nearsightedness property
to accelerate the FOE calculation.
Since the slab phase is uniform in the transverse directions
and the $\xi$ dependence of the density is analytically obtained,
we use the nearsightedness in the $z$ direction.
In Fig.~\ref{fig:rho_kappa_rn}, we show the normal and the pair densities 
at $T=0.1$ MeV computed by assuming the vanishing off-diagonal elements, $\rho_{ij}=0$ and $\kappa_{ij}=0$ at $|z_i-z_j|>r_t$.
The truncation distance is chosen as $r_t = 3.33, 4.99,
16.65\unit{fm}$, and $\infty$ (no truncation).
The result with $r_t=4.99\unit{fm}$ qualitatively agrees with
that without the truncation ($r_t=\infty$),
though a small oscillation is seen.
Those of $r_t=16.65\unit{fm}$ and $r_t=\infty$
are indistinguishable.
Therefore, it is possible to adopt $r_t$
significantly smaller than the size of the system
to save computation time.
At higher temperature,
the nearsightedness is even stronger \cite{Nakatsukasa:2022hpp}.
This may not be so advantageous for the present 1D slab phase,
but may be significant for the 2D and 3D systems.


\section{Conclusion}
\label{sec:concl}

We generalize the finite-temperature coordinate-space FOE method of KS theory to that of HFB band theory.
We have proven
a crucial identity that connects the generalized density matrix
and the Fermi-Dirac distribution function
of the HFB Hamiltonian,
i.e. $R^{\bm{k}}= e^{i\bm{k}\cdot\hat{\bm{x}}}f(\tilde{H}_{\rm HFB}^{\bm{k}}) 
e^{-i\bm{k}\cdot\hat{\bm{x}}} $.
The FOE calculates
$f(\tilde{H}_{\text{HFB}}^{\bm{k}})$, expanded
into a series of Chebyshev polynomials.
We obtain the normal and pair densities in a slab phase without diagonalizing the HFB Hamiltonian,
and show that the FOE produces results with high accuracy.
The FOE method is even more effective for
higher temperature and stronger pairing.
The nearsightedness property exists both in
the normal and pair densities,
which may be used to accelerate the numerical computation.
In conclusion, the FOE is a useful method in the HFB band theory in the coordinate-space representation.
It provides a promising tool for the simulation of the pasta phases in the inner crust of neutron stars.

Further extensions of the present work include but are not limited to
(1) fully self-consistent calculations with
modern nuclear EDFs,
(2) optimizing the shape of slabs,
(3) generalizing to the 2D and 3D phases,
and 
(4) extracting transport properties of free neutrons.

\begin{acknowledgments}
    This research is supported by KAKENHI under Grant No. JP23K25864
    and Japan Science and Technology Agency ERATO Grant No. JPMJER2304.
    We used computational resources provided by the Multidisciplinary Cooperative Research Program
    in Center for Computational Sciences, University of Tsukuba.
\end{acknowledgments}

\bibliographystyle{apsrev4-2}
\bibliography{refs.bib}

\end{document}